# Tuning Single-Atom Electron Spin Resonance in a Vector-Magnetic Field

*Philip Willke[1,2,#], Aparajita Singha[1,2,#], Xue Zhang[1,2,#], Taner Esat[1,2], Christopher P. Lutz[3], Andreas J. Heinrich[1,4,*], Taeyoung Choi[1,4,*]*

1 Center for Quantum Nanoscience, Institute for Basic Science (IBS), Seoul 03760, Republic of Korea
2 Ewha Womans University, Seoul 03760, Republic of Korea
3 IBM Almaden Research Center, San Jose, CA, USA
4 Department of Physics, Ewha Womans University, Seoul 03760, Republic of Korea
# These authors contributed equally to this work.

*E-mail: corresponding author: choi.taeyoung@qns.science, heinrich.andreas@qns.science

**Spin resonance of single spin centers bears great potential for chemical structure analysis, quantum sensing and quantum coherent manipulation. Essential for these experiments is the presence of a two-level spin system whose energy splitting can be chosen by applying a magnetic field. In recent years, a combination of electron spin resonance (ESR) and scanning tunneling microscopy (STM) has been demonstrated as a technique to detect magnetic properties of single atoms on surfaces and to achieve sub-μeV energy resolution. Nevertheless, up to now the role of the required magnetic fields has not been elucidated. Here, we perform single-atom ESR on individual Fe atoms adsorbed on magnesium oxide (MgO), using a 2D vector magnetic field as well as the local field of the magnetic STM tip in a commercially available STM. We show how the ESR amplitude can be greatly improved by optimizing the magnetic fields, revealing in particular an enhanced signal at large in-**

**plane magnetic fields. Moreover, we demonstrate that the stray field from the magnetic STM tip is a versatile tool. We use it here to drive the electron spin more efficiently and to perform ESR measurements at constant frequency by employing tip-field sweeps. Lastly, we show that it is possible to perform ESR using only the tip field, under zero external magnetic field, which promises to make this technique available in many existing STM systems.**

The combination of scanning probe methods and coherent control of spin systems has provided unique access to quantum systems, for instance using NV-centers or magnetic resonance force microscopy[1,2] enabling complex magnetic structures to be visualized[3-6]. In recent years, the combination of electron spin resonance (ESR) and scanning tunneling microscopy (STM) has entered as another platform for controlling atomic spins on surfaces[7]. This technique permits the study of magnetic interaction between pairs of atoms with unprecedented resolution[8,9] and the investigation of the phase coherent properties of atomic spins[10,11]. It even allows access to the hyperfine interaction of single atoms[12,13]. It has spiked strong interest in theoretical studies concerning the phase coherence as well as the driving mechanism of the surface atom spins[7,14-20]. Despite its potential applications, this technique has barely been adopted so far[21,22]. In the case of single Fe atom ESR experiments, measurements were thus far only conducted in a magnetic field with a strong in-plane component, at temperatures at or below 4 K[10,22] and with cabling that has high transmission in the radio frequency (RF) range[21-23].

In this letter, we explore the evolution of the resonance frequency and optimize the amplitude of the ESR signal by using an adjustable 2D vector magnetic field. We show that the strong in-plane magnetic field used previously is not essential. In addition, we show how to utilize the magnetic field from the spin-polarized STM tip (SP tip)[24-26] to assist the external field by adjusting the Zeeman-splitting and driving the ESR. Ultimately, we show that it is possible to perform single-atom ESR by using the tip field only. This eliminates the need of applying an external magnetic field. Furthermore, it relaxes the requirement for high-frequency cabling having good transmission over a wide range of frequencies, because it can operate at a single fixed radio frequency.

Experiments were conducted in a commercial STM system (Unisoku USM1300) with a maximum external vector magnetic field of $B_z^{ext} = 6$ T and $B_\parallel^{ext} = 5$ T (2 T in vector operation, STM temperature $T = 0.3 – 1$ K). The experimental setup is depicted in Fig. 1a. The measurements were performed on well-isolated individual Fe atoms adsorbed atop two atomic layers of MgO grown on a Ag(100) substrate (Fig. 1b). We here follow the ESR implementation described in Ref. (7,23) (See supporting information for further details). For ESR-STM measurements, an RF voltage $V_{RF}$ is added to the DC bias voltage $V_{DC}$ and the radio frequency $f$ is swept. This RF voltage drives transitions between the two lowest-lying Zeeman-split states of the Fe atom, and the change of state population is detected by a difference in tunneling current $\Delta I$ through tunneling magneto-resistance[7]. The Fe atoms on this surface have strong out-of-plane ($z$-axis) anisotropy, so their resonance frequency is given by

$$f_0 = 2\mu_{\text{Fe}}/h \cdot [B_z^{\text{ext}} + B_z^{\text{tip}}] \quad (1)$$

where $\mu_{\text{Fe}}$ is the magnetic moment of Fe, $h$ is Planck's constant and $B_z^{\text{ext}}$ is the external magnetic field in z-direction. $B_z^{\text{tip}}$ is the z-axis component of the magnetic field generated by the SP tip[24-26], which is made magnetic by picking up several Fe atoms (typically 1-3) from the surface.

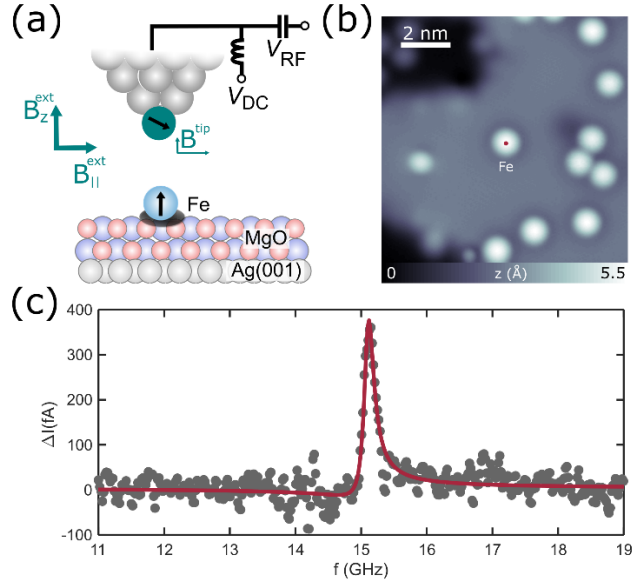

**Figure 1**. Electron spin resonance in a scanning tunneling microscope with a vector magnet. (a) Schematic of the experiment. Single Fe atoms on bilayer MgO grown on Ag(001) are studied using ESR-STM. An external vector magnet allows to adjust the field in-plane $B_\parallel^{ext}$ and out-plane $B_z^{ext}$ in addition to the field $B^{tip}$ created by the spin-polarized magnetic tip. In addition to the DC bias voltage $V_{DC}$ a radio frequency (RF) voltage $V_{RF}$ at frequency $f$ is applied to the tunnel junction. (b) Constant-current topography showing several single atoms on an MgO surface, the central one being a Fe atom ($I = 50$ pA, $V_{DC} = 60$ mV, $T = 0.9$ K), and (c) ESR spectrum taken on the center of the Fe atom (red dot). Red line is a fitted Fano-Lorentzian curve. ($I = 10$ pA, $V_{DC} = 40$ mV, $V_{RF} = 10$ mV, $T = 0.9$ K, $B_\parallel^{ext} = 1.4$ T, $B_z^{ext} = 0.12$ T ).

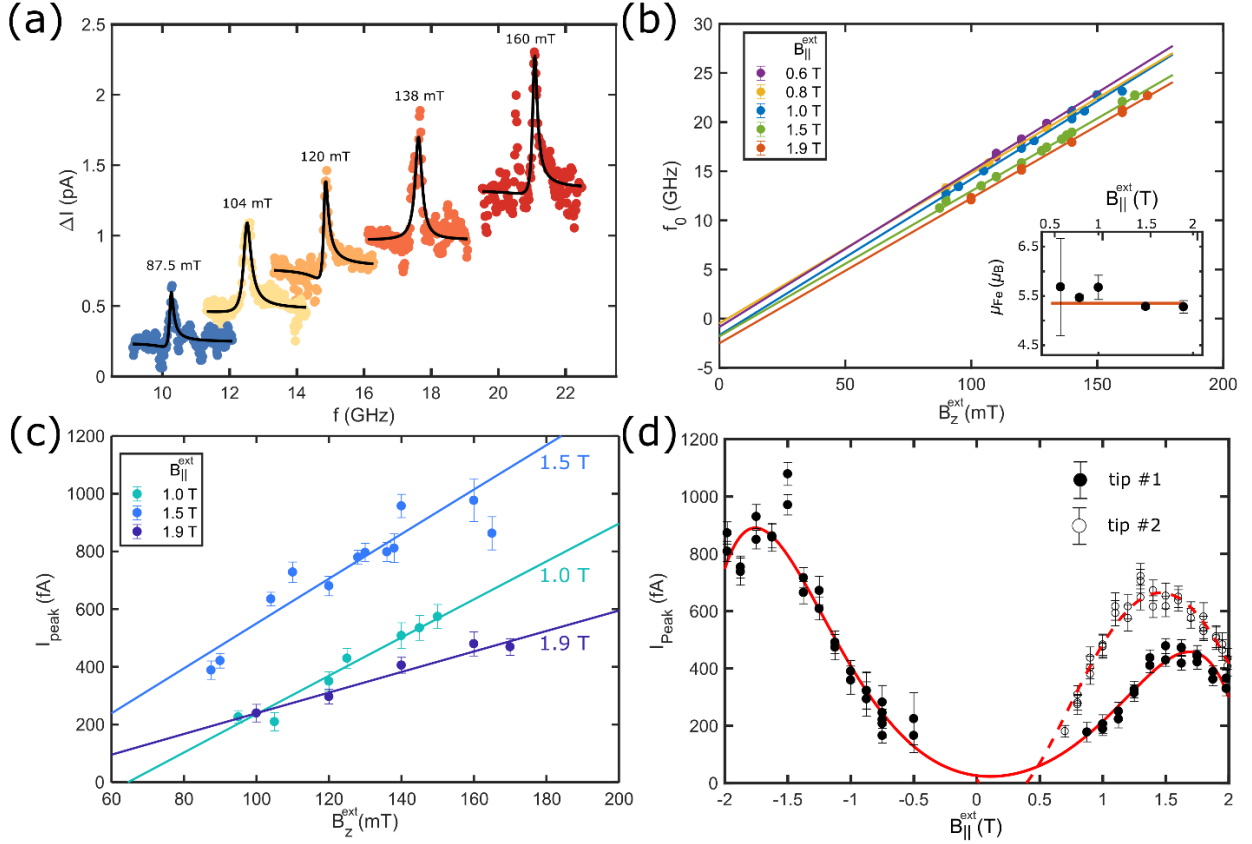

**Figure 2.** Magnetic-field dependence of single-atom ESR peaks. (a) ESR spectra taken on the Fe atom in Fig. 1 for different external magnetic field $B_z^{\text{ext}}$. Plots are shifted vertically for clarity ($I = 50$ pA, $V_{\text{DC}} = 60$ mV, $V_{\text{RF}} = 22$ mV, $B_{\parallel}^{\text{ext}} = 1.5$ T, $T = 0.8$ K. Additional peak at 160 mT stems from a bad RF transmission at that frequency.). (b) Resonance frequencies $f_0$ as a function of $B_z^{\text{ext}}$ for different $B_{\parallel}^{\text{ext}}$. Influence of the tip-magnetic field has been subtracted for all curves (See supporting information). Inset: Magnetic moment of Fe $\mu_{\text{Fe}}$, extracted from the slopes for each $B_{\parallel}^{\text{ext}}$ [Eq. (1)]. Red line shows the mean value. (c) ESR peak amplitude $I_{\text{peak}}$ as a function of $B_z^{\text{ext}}$ for three $B_{\parallel}^{\text{ext}}$ and (d) as a function of $B_{\parallel}^{\text{ext}}$. In panel (d), all measurements were conducted at $f_0 = 17 \pm 0.5$ GHz [red line: guide to the eye, $I = 50$ pA, $V_{\text{DC}} = 60$ mV, $V_{\text{RF}} = 15$ mV (tip#1), 22 mV (tip#2), $T = 0.9$ K]. Closed (open) circles are taken for tip#1 (tip#2).

Various theoretical proposals have addressed the mechanism of single-atom ESR[7,14-20]. Recent experiments, performed on Ti atoms on the surface, suggest that the atom is exposed to a time-varying tip magnetic field component perpendicular to the Zeeman field, similar to conventional ESR. This mechanism works because the atom is mechanically shaken on the polar surface of MgO by the oscillating $V_{RF}$. Consequently, the Rabi rate $\Omega$, characterizing how fast a spin can be driven coherently, increases linearly with the tip magnetic field gradient[26].

Figure 1c shows a typical ESR peak taken on a single Fe atom. The shape of the resonance is Lorentzian $\Delta I = I_{peak} \cdot [1 + \delta^2]^{-1}$, given by the steady-state solution of the Bloch equations[7] (See Ref. [11] for a detailed treatment of the Fano-shaped asymmetric contribution). Here, $\delta = 2(f - f_0)/\Gamma$ with $\Gamma$ being the linewidth of the ESR peak. The ESR peak amplitude $I_{peak}$ is determined by several experimental parameters[10]

$$I_{peak} = I \cdot 2\eta \cdot [P_0 - \tfrac{1}{2}] \cdot \Phi(\Omega) \qquad (2)$$

Here, $I$ is the DC tunneling current applied during an ESR sweep, $\eta$ is the spin polarization of the tip and $P_0$ is the ground state population probability of the Fe spin when off-resonance. The driving factor $\Phi(\Omega) = T_1 T_2 \Omega^2/(1 + T_1 T_2 \Omega^2)$ ranges from 0 to 1 and characterizes how much the spin population is driven into equal state population. It depends on the Rabi rate $\Omega$, the spin relaxation time $T_1$ and phase coherence time $T_2$ (See supporting information).

In the following, we make use of our 2D external magnetic field in order to analyze the change in the properties of the ESR signal. We measure ESR at tunneling conductances ($\sigma = 50~\text{pA}/60~\text{mV} = 0.8~\text{nS}$) that are comparable to those used in Ref. [7,10]. For these tunneling conditions, we find that the z-axis component of the magnetic field from the SP tip ($B_z^{tip}$) is less than 10% of $B_z^{ext}$. Figure 2a shows ESR measurements as a function of $B_z^{ext}$ at a fixed $B_{||}^{ext} = 1.5$ T. We find that $f_0$ shifts linearly with $B_z^{ext}$, as described by Eq. (1). This linear relation holds when the experiment is repeated for different $B_{||}^{ext}$ as shown in Fig. 2b. Given a strong out-of-plane magnetic anisotropy of Fe[27], its magnetic moment does not align with the in-plane magnetic field. This explains the unchanged slope for varying $B_{||}^{ext}$ and results in a constant magnetic moment of $\mu_{Fe} = (5.35 \pm 0.14)~\mu_B$ (Fig. 2b, inset). This is in excellent agreement with previous works[7,8,27]. The lines deviate from the origin ($B_z^{ext} = 0$ and $f_0 = 0$) which is caused by a small misalignment [∼(1–2)°] between the sample and the magnetic field axes. This misalignment leads to an additional $B_z^{ext}$ component when applying $B_\parallel^{ext}$. Measurements on different atoms revealed only small variations of ~3 % in $f_0$, in contrast to earlier reports[7] (see supplemental information).

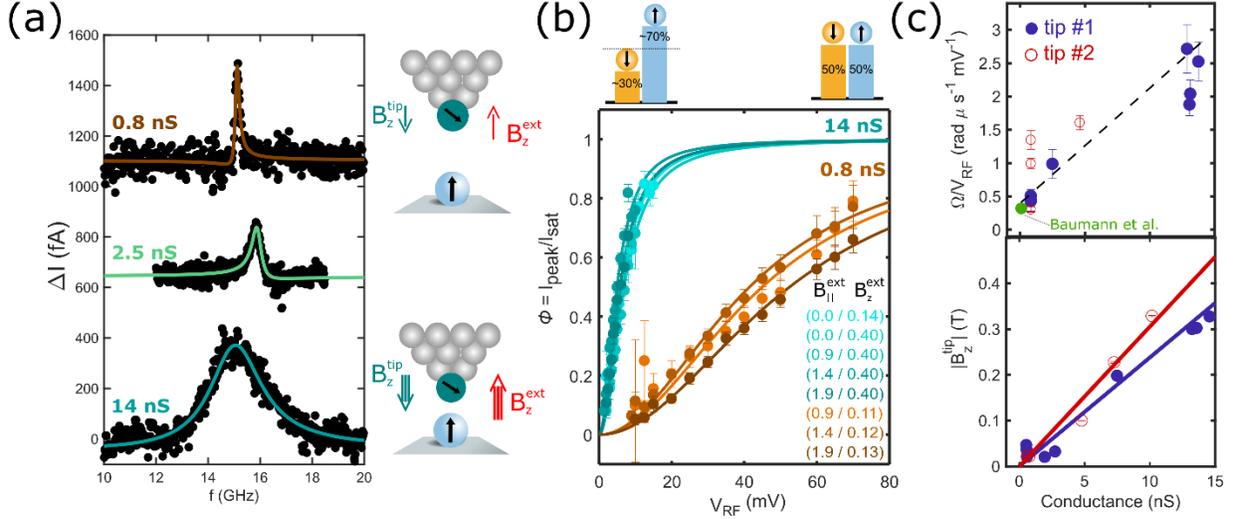

**Figure 3**. Improvement of the Rabi rate by increasing the tip field $B^{tip}$. (a) ESR spectra taken for different tunneling conductances and varying external magnetic field, illustrating that ESR at zero in-plane field is made possible by using higher conductance. (Tunneling parameters: top: $I = 50$ pA, $V_{DC} = 60$ mV, $V_{RF} = 10$ mV, $B_\parallel^{ext} = 1.4$ T, $B_z^{ext} = 0.12$ T / middle: $I = 20$ pA, $V_{DC} = -8$ mV, $V_{RF} = 8$ mV, $B_\parallel^{ext} = 0$ T, $B_z^{ext} = 0.135$ T / bottom: $I = 115$ pA, $V_{DC} = 8$ mV, $V_{RF} = 3$ mV, $B_\parallel^{ext} = 0$ T, $B_z^{ext} = 0.4$ T. $T = 1$ K for all spectra. Curves have been offset vertically for clarity). (b) Saturation measurements $\Phi(V_{RF}) = I_{peak}(V_{RF})/I_{sat}$ which give the ESR peak amplitude $I_{peak}$ normalized to $I_{sat} = I \cdot 2\eta \cdot [P_0 - \frac{1}{2}]$ (see Eq. 2 and Ref. (7)). $I_{sat}$ is the maximum value of $I_{peak}$ in the limit of large $V_{RF}$, obtained when the spin system is saturated (driven to a 50%/50% state population [$\Phi(\Omega) = 1$] as indicated in the sketch above). Colors indicate different external magnetic field settings as specified in the inset (Units in [T]). (c) Top: Rabi rate $\Omega$ normalized by the RF voltage ($V_{RF}$) for different tunneling conductances $\sigma = I/V_{DC}$. Green dot marks the value obtained in Ref. (7,10). Closed (open) circles are for tip #1 (#2). Dashed line is a guide to the eye. Bottom: z-component of the tip field $B_z^{tip}(\sigma)$ (see supporting information). The solid lines are linear fits to the data.

With the benefit of a 2D vector magnet, we now investigate the ESR peak amplitude $I_{peak}$ as a function of $B_z^{ext}$ and $B_\parallel^{ext}$. Figure 2a indicates that $I_{peak}$ increases with $B_z^{ext}$. The peak amplitudes are plotted in Fig. 2c for three different $B_{\parallel}^{\text{ext}}$. The increase with $B_z^{\text{ext}}$ for a given $B_{\parallel}^{\text{ext}}$ is mainly caused by an improved ground state population $P_0$ [Eq. (2)], since the polarization into the ground state approximately follows $\left[P_0 - \frac{1}{2}\right] \propto B_z^{\text{ext}}$ (See supporting information). However, the peak amplitudes at different $B_{\parallel}^{\text{ext}}$ are strikingly different, which is shown in Fig. 2D in greater detail. Overall, $I_{\text{peak}}$ increases with in-plane magnetic field and is maximized at $B_{\parallel}^{\text{ext}} \approx \pm 1.5$ T.

For now, we cannot attribute the trend of the ESR peak amplitude to one dominant effect and it is likely that several factors are contributing. The monotonic increase of the ESR peak amplitude up to ~1.5 T can be explained by an increased SP tip polarization $\eta$ and an improved driving term $\Phi(\Omega)$ as evident from Eq. (2). Indeed, theoretical calculations suggest that a higher $B_{\parallel}^{\text{ext}}$ results in a higher Rabi rate $\Omega$ and therefore higher $\Phi(\Omega)$[7,14]. At $B_{\parallel}^{\text{ext}}$ exceeding ~1.5 T, we find that the ESR peak amplitude decreases, which might be associated with a decrease of spin relaxation time $T_1$, a trend that we find in pump-probe spectroscopy measurements (supplementary material).

We have not observed any ESR-signal for $B_{\parallel}^{\text{ext}} < 0.5$ T at small tunneling conductance $\sigma$, in the range of 0.8 nS. This may be associated either with the tip's spin polarization being too low, or a strongly diminished Rabi rate. Interestingly, we successfully recover the ESR signal by increasing the tunneling conductance $\sigma$ in the STM junction as shown in Fig 3a. As the STM tip gets closer to the Fe atom, we find that its stray magnetic field $B_z^{\text{tip}}$ shifts the resonance of the Fe atom to lower frequencies. Subsequently, we compensate the shift of $f_0$ by increasing $B_z^{\text{ext}}$ and thus maintain the resonance frequency constant (~15 GHz). In the case of higher conductance (stronger tip field), we obtain an ESR signal even when lowering $B_{\parallel}^{\text{ext}}$ down to 0 T (Fig. 3a). This suggests that the increased tip magnetic field facilitates driving ESR under these conditions.

As the conductance increases, the linewidth of the resonance becomes broader as well. This is likely caused by a decrease in $T_2$[10], as well as mechanical vibrations of the tip, which lead to fluctuations in $B_z^{\text{tip}}$[7]. Nevertheless, we find that for higher conductances ESR can be readily obtained even though the RF voltage is low. This is manifested in Fig. 3b, showing the driving factor $\Phi(\Omega)$ defined in Eq. (2) as a function of $V_{\text{RF}}$ for various conductances and $B^{\text{ext}}$ field configurations. Accordingly, we find that the ESR signal saturates at much lower RF voltages $V_{\text{RF}}$ as the conductance is increased (Fig. 3b, see supporting information and Ref. [10]). The saturation curves show negligible dependence on $B_{\parallel}^{\text{ext}}$, indicating that the tip field dominates the saturation.

To obtain the Rabi rate $\Omega$ from these measurements, $T_1$ and $T_2$ need to be determined under the respective tunneling parameters. Here, we estimate them using $T_1$ measurements under different parameters in combination with relations for $T_1$ and $T_2$ found previously[10,28](supporting information).

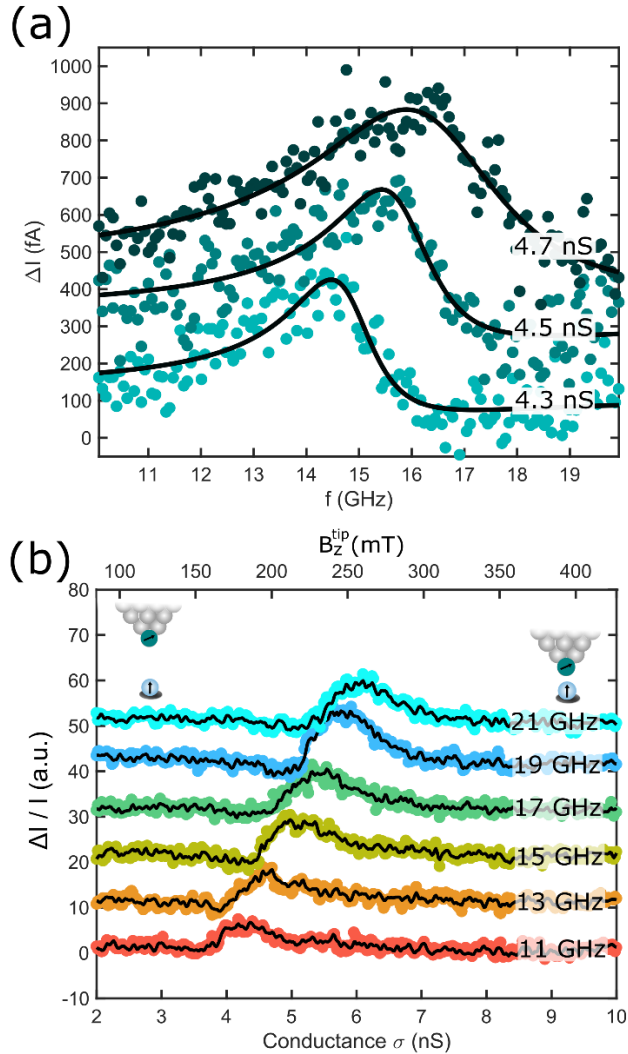

**Figure 4**. ESR without external magnetic field. (a) ESR spectra taken at zero external magnetic field $B_\parallel^{ext} = B_z^{ext} = 0$ T for different conductances $\sigma$ ($V_{DC} = 30$ mV, $V_{RF} = 15$ mV, $T = 0.4$ K). Spectra shift in frequency due to the increasing magnetic field from the tip. Vertical offsets are added for clarity. (b) ESR spectrum for different fixed frequencies measured by sweeping the conductance $\sigma$ ($I = 100 - 500$ pA, $V_{DC} = 50$ mV, $V_{RF} = 25$ mV). This effectively increases the tip field $B_z^{tip} \propto \sigma$ (top scale). ESR signal $\Delta I$ is normalized to compensate for the linear increase with $I$ [See Eq. (2)]. Vertical offsets have been added for clarity and a nonlinear background was subtracted (see supporting information).

The extracted $\Omega$ as a function of $\sigma$ is shown in Fig. 3c. Since $\Omega$ scales linearly with $V_{RF}$[7,10,14], we normalize the Rabi rate $\Omega/V_{RF}$ to obtain a measure of the driving efficiency. In addition, we plot the change in tip field $B_z^{tip}$ projected onto the z-axis as a function of $\sigma$ (evaluation see supporting information).

The two proportionalities $\Omega \propto \sigma$ and $B_z^{tip} \propto \sigma$ found here imply $\Omega \propto B_z^{tip}$. However, other measurements and theoretical models suggest an improvement in $\Omega$ for an increasing magnetic field gradient $\partial B_z^{tip}/\partial z$[14,26]. Still, this can be easily fulfilled in case of magnetic exchange coupling, because of the exponential distance dependence that is proportional to its own derivative (supporting information).

These results indicate that a magnetic tip field can drive single-atom ESR. However, we cannot fully exclude the possibility of a crystal field driving mechanism as originally proposed in Ref. (7), since the electric field in the tunnel junction also increases for higher conductance.

Nevertheless, we achieve Rabi rates $\Omega$ on Fe atoms approximately one order of magnitude higher than previous experiments[7,10] (Fig. 3c) by utilizing the proximity of the tip. We note that in the case of spin-1/2 Ti atoms, comparable $\Omega$ have been obtained[26]. In that case, the absence of an anisotropy barrier facilitates driving of Ti atoms, but at the same time limits their $T_1$ to ~100 ns[9]. In contrast, Fe atoms revealed a $T_1$ that is three orders of magnitude higher[7,28].

Lastly, we demonstrate ESR on individual Fe atoms without any external magnetic field $B_\parallel^{ext} = B_z^{ext} = 0$ T by employing the tip field to give the Zeeman splitting (Fig. 4a). The resonance frequency shifts up for increasing conductance $\sigma$, since $B_z^{tip} \propto \sigma$. The ESR peak is even broader than those in Fig 3a, likely due to current-induced relaxation and decoherence[10,28]. As a consequence, our frequency range is now too narrow to reveal the full ESR peak. Therefore, we ultimately switch from sweeping $f$ to sweeping the conductance $\sigma$ at a fixed frequency while the current feedback loop remains enabled. Thus, the tip approaches the atom establishing a sweep in $B_z^{tip}$. This is demonstrated in Fig. 4b where we perform tip-field ESR sweeps. As expected, the resonance linearly shifts to higher tunneling conductances for increasing setpoint frequency [Eq. (1)]. In contrast to the broad ESR signal as a function of frequency (Fig. 4a), this method offers a significantly larger range of ~300 mT tip magnetic field which would capture a frequency window of equivalently ~40 GHz. It also allows for a considerable speed-up of the measurement, since the tip magnetic field can be changed much faster (>1 kHz) than the frequency of the RF generator (<10 Hz).

Our work illustrates how to tune the ESR properties by combining vector magnet and tip field and eventually realize single-atom ESR with tip field only. This suggests a way to study nanoscale magnetic systems revealing for instance magnetic bistability[29,30] and quantum tunneling of magnetization[31] in the absence of an external field. Moreover, demonstrating ESR-STM in a commercially available STM will allow this technique to be performed in a greater class of STM systems, even those without an external magnetic field.

ASSOCIATED CONTENT

**Supporting Information**. The supplementary information covers sample preparation and experimental methods, ESR measurements on different atoms, pump-probe spectroscopy measurements, discussion on the peak height intensity, estimation of the Rabi rate, evaluation of the tip magnetic field as a function of conductance, discussion on the relation between Rabi rate and tip field and the evaluation and supplemental data on the ESR without macroscopic field.

AUTHOR INFORMATION

Corresponding Author

*E-Mail: choi.taeyoung@qns.science,

*E-Mail: heinrich.andreas@qns.science

**Author Contributions**

P.W., A.S., X.Z., T.E., and T.C. performed the experiments. T.C. and A.J.H. supervised the project. P.W. wrote the manuscript with contributions from all authors. All authors discussed the results. # P.W., A.S. and X.Z. contributed equally to this work.

**Notes**

The authors declare no competing financial interest.

ACKNOWLEDGMENT

All authors acknowledge support from the Institute for Basic Science under grant IBS-R027-D1. P.W. acknowledges support from the Alexander von Humboldt Foundation. We acknowledge Minhee Choi for fruitful discussions. We also thank William Paul for providing software to control the RF source.

REFERENCES

1. Balasubramanian, G.; Chan, I. Y.; Kolesov, R.; Al-Hmoud, M.; Tisler, J.; Shin, C.; Kim, C.; Wojcik, A.; Hemmer, P. R.; Krueger, A.; Hanke, T.; Leitenstorfer, A.; Bratschitsch, R.; Jelezko, F.; Wrachtrup, J. Nanoscale imaging magnetometry with diamond spins under ambient conditions. *Nature* **2008**, 455, 648-651.

2. Rugar, D.; Budakian, R.; Mamin, H.J.; Chui, B.W. Single spin detection by magnetic resonance force microscopy. *Nature* **2004**, 430, 329-332.

3. Degen, C.L.; Poggio, M.; Mamin, H.J.; Rettner, C.T.; Rugar, D.; Nanoscale magnetic resonance imaging. *PNAS* **2009**, 106, 1313-1317.

4. Gross, I.; Akhtar, W.; Garcia, V.; Martínez, L.J.; Chouaieb, S.; Garcia, K.; Carrétéro, C.; Barthélémy, A.; Appel, P.; Maletinsky, P.; Kim, J.V.; Chauleau, J. Y.; Jaouen, N. ; Viret, M.; Bibes, M.; Fusil, S.; Jacques, V. Real-space imaging of non-collinear antiferromagnetic order with a single-spin magnetometer. *Nature* **549**, 252-256 (2017).

5. Casola, F.; van der Sar, T.; Yacoby, A.; Probing condensed matter physics with magnetometry based on nitrogen-vacancy centres in diamond. *Nat. Rev. Mater.* **2018**, 3, 17088.

6. Thiel, L.; Wang, Z.; Tschudin, M.A.; Rohner, D.; Gutiérrez-Lezama, I.; Ubrig, N.; Gibertini, M.; Giannini, E.; Morpurgo, A.F.; Maletinsky, P.; Probing magnetism in 2D materials at the nanoscale with single spin microscopy. *Science* **2019,** 364, 973-976.

7. Baumann, S.; Paul, W.; Choi, T.; Lutz, C.P.; Ardavan, A.; Heinrich, A.J. Electron paramagnetic resonance of individual atoms on a surface. *Science* **2015**, 350, 417-420.

8. Choi, T.; Paul, W.; Rolf-Pissarczyk, S.; Macdonald, A.J.; Natterer, F.D.; Yang, K.; Willke, P.; Lutz, C.P.; Heinrich, A.J. Atomic-scale sensing of the magnetic dipolar field from single atoms. *Nature Nano.* **2017**, 12(5), 420-424.

9. Yang, K.; Bae, Y.; Paul, W.; Natterer, F.D.; Willke, P.; Lado, J.L.; Ferrón, A.; Choi, T.; Fernández-Rossier, J.; Heinrich, A.J.; Lutz, C.P. Engineering the eigenstates of coupled spin-1/2 atoms on a surface. *Phys. Rev. Lett.* **2017**, 119, 227206.

10. Willke, P.; Paul, W.; Natterer, F.D.; Yang, K.; Bae, Y.; Choi, T.; Fernández-Rossier, J.; Heinrich, A.J.; Lutz, C.P. Probing quantum coherence in single-atom electron spin resonance. *Science Adv.* **2018**, 4(2), eaaq1543.

11. Bae, Y.; Yang, K.; Willke, P.; Choi, T.; Heinrich, A.J.; Lutz, C.P. Enhanced quantum coherence in exchange coupled spins via singlet-triplet transitions. *Science Adv.* **2018**, 4(11), eaau4159.

12. Willke, P.; Bae, Y.; Yang, K.; Lado, J.L.; Ferrón, A.; Choi, T.; Ardavan, A.; Fernández-Rossier, J.; Heinrich, A.J.; Lutz, C.P. Hyperfine interaction of individual atoms on a surface. *Science* **2018**, 362(6412), 336-339.

13. Yang, K.; Willke, P.; Bae, Y.; Ferrón, A.; Lado, J.L.; Ardavan, A.; Fernández-Rossier, J.; Heinrich, A.J.; Lutz, C.P. Electrically controlled nuclear polarization of individual atoms. *Nature Nano*. **2018**, 13(12), 1120-1125.

14. Lado, J.L.; Ferrón, A.; Fernández-Rossier, J. Exchange mechanism for electron paramagnetic resonance of individual adatoms. *Phys. Rev. B* **2017**, 96(20), 205420.

15. Shakirov, A.M.; Rubtsov, A.N.; and Ribeiro, P. Spin transfer torque induced paramagnetic resonance. *Phys. Rev. B* **2019**, 99(5), 054434.

16. Berggren, P.; Fransson, J. Electron paramagnetic resonance of single magnetic moment on a surface. *Scientific reports* **2016**, 6, 25584.

17. Shavit, G.; Horovitz, B.; Goldstein, M. Generalized open quantum system approach for the electron paramagnetic resonance of magnetic atoms. *Phys. Rev. B* **2019**, 99(19), 195433.

18. Delgado, F.; Fernández-Rossier, J. Spin decoherence of magnetic atoms on surfaces. *Prog. Surf. Sci.* **2017**, 92(1), 40-82.

19. Ibañez-Azpiroz, J.; dos Santos Dias, M.; Blügel, S.; Lounis, S. Longitudinal and transverse spin relaxation times of magnetic single adatoms: An ab initio analysis. *Phys. Rev. B* **2017**, 96(14), 144410.

20. Gálvez, J.R.; Wolf, C.; Delgado, F.; Lorente, N. Cotunneling mechanism for all-electrical electron spin resonance of single adsorbed atoms. *Phys. Rev. B* **2019**, 100(3), 035411.

21. Natterer, F.D.; Patthey, F.; Bilgeri, T.; Forrester, P.R.; Weiss, N.; Brune, H. Upgrade of a low-temperature scanning tunneling microscope for electron-spin resonance. *Rev. Sci. Instrum.* **2019**, 90(1), 013706.

22. Seifert, T.S.; Kovarik, S.; Nistor, C.; Persichetti, L.; Stepanow, S.; Gambardella, P. Single-atom electron paramagnetic resonance in a scanning tunneling microscope driven by a radiofrequency antenna at 4 K. *arXiv preprint* **2019**, arXiv:1908.03379.

23. Paul, W.; Baumann, S.; Lutz, C. P.; Heinrich, A. J. Generation of constant-amplitude radio-frequency sweeps at a tunnel junction for spin resonance STM. *Rev. Sci. Instrum.* **2016**, 87, 074703.

24. Yan, S.; Choi, D.J.; Burgess, J.A.; Rolf-Pissarczyk, S.; Loth, S. Control of quantum magnets by atomic exchange bias. *Nature Nano.* **2010**, 10(1), 40-45.

25. Willke, P.; Yang, K.; Bae, Y.; Heinrich, A. J.; Lutz, C. P. Magnetic Resonance Imaging of Single Atoms, *Nat. Phys.* **2019**.

26. Yang, K.; Paul, W.; Natterer, F.D.; Lado, J.L.; Bae, Y.; Willke, P.; Choi, T.; Ferrón, A.; Fernández-Rossier, J.; Heinrich, A.J.; Lutz, C.P. Tuning the Exchange Bias on a Single Atom from 1 mT to 10 T. *Phys. Rev. Lett.* **2019**, 122(22), 227203.

27. Baumann, S.; Donati, F.; Stepanow, S.; Rusponi, S.; Paul, W.; Gangopadhyay, S.; Rau, I.G.; Pacchioni, G.E., Gragnaniello, L., Pivetta, M.; Dreiser, J.; Piamonteze, C.; Lutz, C. P.; Macfarlane, R. M.; Jones, B. A.; Gambardella, P.; Heinrich, A. J.; Brune, H. Origin of perpendicular magnetic anisotropy and large orbital moment in Fe atoms on MgO. *Phys. Rev. Lett.* **2015**, 115(23), 237202.

28. Paul, W.; Yang, K.; Baumann, S.; Romming, N.; Choi, T.; Lutz, C.P.; Heinrich, A.J. Control of the millisecond spin lifetime of an electrically probed atom. *Nat. Phys.* **2017**, 13, 403-407.

29. Khajetoorians, A.A.; Baxevanis, B.; Hübner, C.; Schlenk, T.; Krause, S.; Wehling, T.O.; Lounis, S.; Lichtenstein, A.; Pfannkuche, D.; Wiebe, J.; Wiesendanger, R. Current-driven spin dynamics of artificially constructed quantum magnets. *Science* **2013**, 339(6115), 55-59.

30. Loth, S.; Baumann, S.; Lutz, C.P.; Eigler, D.M.; Heinrich, A.J. Bistability in atomic-scale antiferromagnets. *Science* **2012**, 335(6065),196-199.

31. Forrester, P.R.; Patthey, F.; Fernandes, E.; Sblendorio, D.P.; Brune, H.; Natterer, F.D. A quantum pathway to overcome the trilemma of magnetic data storage. *arXiv preprint* **2019**, arXiv:1903.00242.

Supporting Information

# Tuning Single-Atom Electron Spin Resonance in a Vector-Magnetic Field

*Philip Willke[1,2,#], Aparajita Singha[1,2,#], Xue Zhang[1,2,#], Taner Esat[1,2], Christopher P. Lutz[3], Andreas J. Heinrich[1,4,\*], Taeyoung Choi[1,4,\*]*

1 Center for Quantum Nanoscience, Institute for Basic Science (IBS), Seoul 03760, Republic of Korea
2 Ewha Womans University, Seoul 03760, Republic of Korea
3 IBM Almaden Research Center, San Jose, CA, USA
4 Department of Physics, Ewha Womans University, Seoul 03760, Republic of Korea
# These authors contributed equally to this work.

*E-mail: corresponding author: choi.taeyoung@qns.science, heinrich.andreas@qns.science

**Table of Contents**

# 1. Sample preparation and experimental methods

*1.1 Sample preparation*

For sample preparation, an Ag(001) single crystal was cleaned by 2 cycles of sputtering and annealing. Then, it was heated to ~700 K and exposed to a Mg flux in an oxygen environment of ~$10^{-6}$ mbar. MgO growth was performed for 12 minutes yielding a MgO coverage of ~2 monolayers. Subsequently, the sample was slowly cooled down to room temperature over a time of 15 min and afterwards transferred in to the cold STM (~4 K). Single Fe atoms were evaporated using e-beam evaporation outside of the STM. To do that the sample manipulator was pre-cooled by touching the sample (~15 min). Next, the sample was quickly transferred into the evaporation chamber where it was exposed to the Fe evaporator for several seconds. For the STM measurements a platinum iridium wire tip was used, which was presumably coated with silver due to indentations into the Ag substrate. The MgO layer thickness was confirmed by point-contact measurements on single Fe atoms[28].

*1.2 Electron spin resonance*

For ESR-measurements we closely followed the schemes described in Refs (7,23). A Keysight E8257D was used to generate the RF voltage which was mixed with the normal DC tunneling bias using a bias tee (SigaTek SB15D2) outside the vacuum chamber. For pump-probe measurements, we used an arbitrary waveform generator (Tektronix 70002A). A National Instrument DAQ-Box 6343 was used to trigger the RF generator and to record the data. The data was taken using a lockin-detection scheme with an on/off modulation of the RF signal at 95 Hz. We here used a SR860 (Stanford Research Systems) lock-in amplifier[23].

ESR experiments were conducted under closed constant-current feedback loop conditions with the integral feedback parameter lowered to ~5 nm/s (Nanonis controller settings).

To obtain frequency-independent RF voltage at the tunnel junction, we measured the RF transfer function using the Fe conductance step in inelastic electron tunneling spectroscopy[23]. The resulting transfer function is shown in Fig. S1 in comparison to that obtained in the Ref. (23).

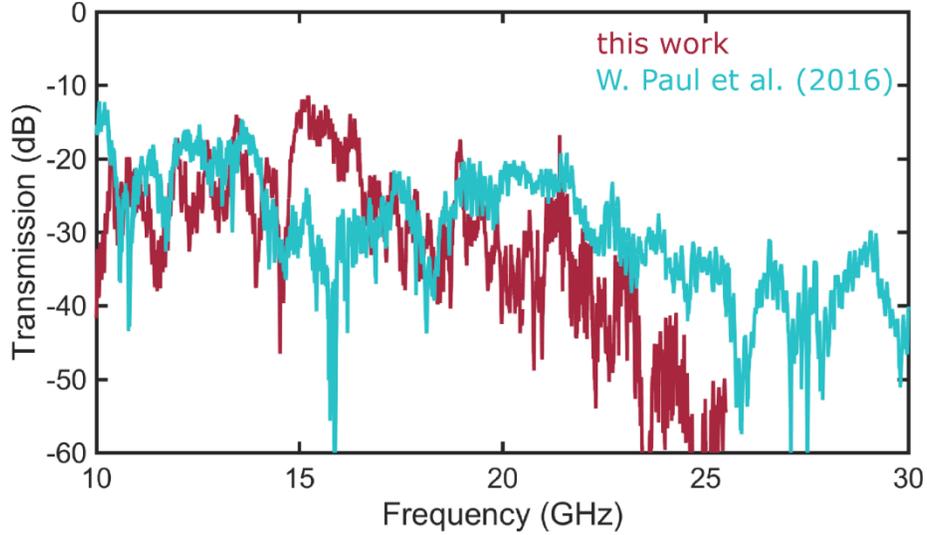

**Figure S1.** Comparison of RF transfer functions showing the transmission from the RF voltage source to the STM tunnel junction. Red: This work. Cyan: Ref. (23).

## 2. ESR measurements on different atoms

In the framework of this study we performed measurements on several Fe atoms. The topographic images of 4 different Fe atoms and their corresponding ESR resonance frequencies $f_0$ as a function of $B_z^{\text{ext}}$ are shown in Fig. S2. For a given $B_z^{\text{ext}}$, we only find minor differences of less than 500 MHz in $f_0$ between the atoms. This is significantly smaller than in previous reports (>3 GHz, Ref. [7]). Previously, the large spread in $f_0$ among atoms was attributed to a subtle different local environment of the atoms leading either to different magnetic moments or to variable tilting of the easy axis relative to the magnetic field. Since these experiments were conducted in a strong in-plane magnetic field ($B_\parallel^{\text{ext}} = 5$ T), a small tilt would indeed greatly change the magnetic field component along the Fe magnetization direction. Our measurements show that the resonance frequency variations are reduced to ~<3 % in the limit of no in-plane magnetic field ($B_\parallel^{\text{ext}} = 0$ T).

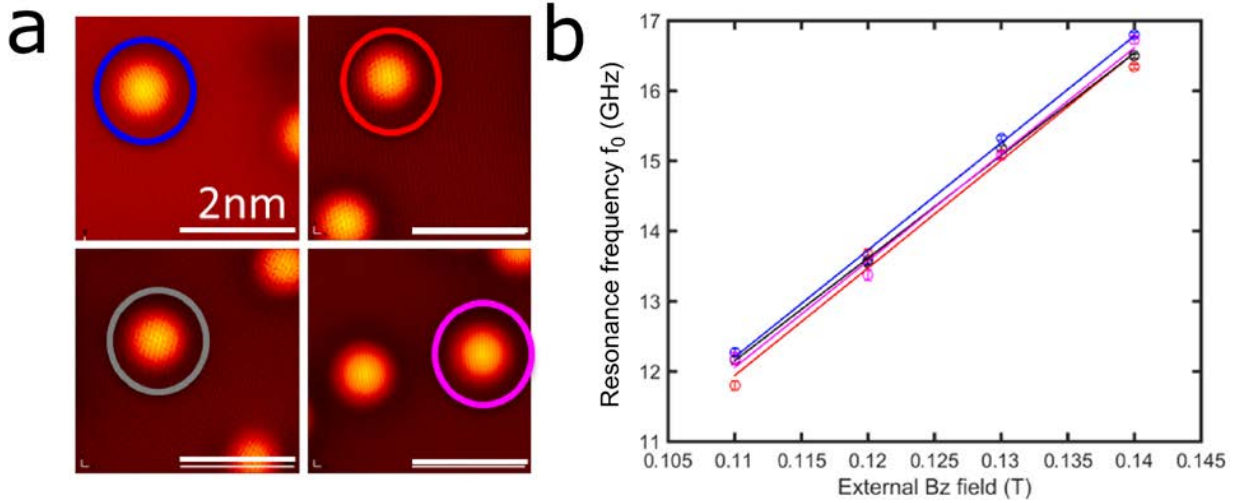

**Figure S2.** Measurements on different atoms. (a) Constant-current topography images of four different Fe atoms ($I = 50$ pA, $V_{DC} = 60$ mV, $T = 1$ K). (b) Resonance frequency $f_0$ of the four atoms as a function of $B_z^{ext}$ field ($I = 20$ pA, $V_{DC} = -8$ mV, $V_{RF} = 10$ mV, $B_\parallel^{ext} = 0$ T, $T = 1$ K). The different colors relate to the colored circles around the atoms in (a).

## 3. Pump-Probe Spectroscopy

To estimate the energy relaxation time $T_1$ of the Fe atom from the excited state $|S_z = -2>$ to the ground state $|S_z = +2>$ at different in-plane magnetic fields $B_\parallel$ we applied an all-electrical pump-probe scheme using a spin-polarized tip[28]. We generated the pump-probe scheme shown in Fig. S3a by using an arbitrary waveform generator (Tektronix 70002A) and applied the scheme to the STM junction (with open-feedback loop). We detected the change of the average spin-polarized tunneling current $\Delta I$ between cycle A and B in the pump-probe scheme (see Fig. S3a) by using lock-in detection. Hence, by varying the delay time $\tau$ between the pump and probe pulse we obtained the time-dependent spin dynamics of the Fe atom. As an example, a pump-probe measurement is shown in Fig. S3b. Fitting an exponential to $\Delta I$ as a function of $\tau$ yields the relaxation time $T_1$.

$T_1$ is required to determine the Rabi rate [See Eq. (S5) and section 5]. However, due to decreasing spin-polarization of the magnetic tip, we were not able to perform pump-probe measurements below $B_z^{ext} < 0.4\ T$, i.e. under our ESR conditions. Instead, we measured $T_1$ at higher $B_z^{ext}$ fields and determined $T_1$ for our ESR conditions by extrapolating the high $B_z$ data. The extracted $T_1$ for different $B_\parallel^{ext}$ at $B_z^{ext} = 0.5$ T are shown in Fig. S3c. Most likely the overall linear decrease of $T_1$ with increasing absolute $B_\parallel^{ext}$ is due to the larger mixing of the $|S_z = -2>$ and $|S_z = +2>$ states for larger in-plane magnetic fields[7]. From previous studies, it is known that $T_1$ of Fe decreases linearly with $B_z^{ext}$ for $B_z^{ext} < 2\,T$[28]. Thus, by measuring $T_1$ at two different $B_z^{ext}$ we estimated the change of $T_1$ with $B_z^{ext}$ to be $\sim 46\,\mu s/T$ for our given tip-sample conductance (10 mV, 50 pA). Knowing this value, we mapped the measured $T_1$ at different $B_\parallel^{ext}$ to our ESR regime, i.e. $B_z^{ext} \approx 0.05 - 0.3\,T$.

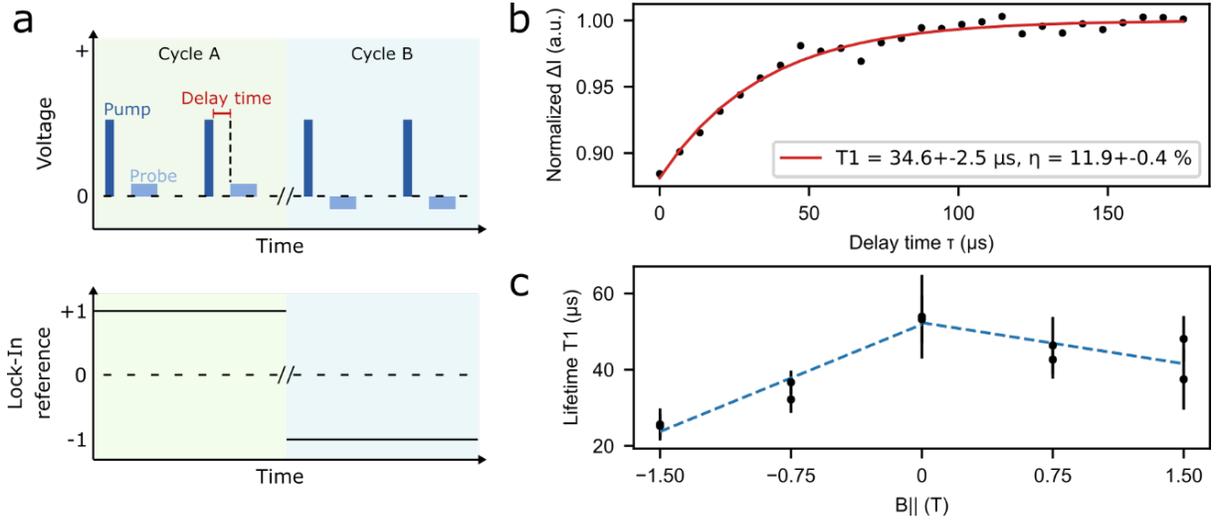

**Figure S3.** Pump-probe measurements on Fe. (a) Illustration of the all-electrical pump-probe measurement scheme (upper panel). By varying the delay time $\tau$ between the pump and probe pulse we obtained the post excitation time dynamics of the Fe spin state by using a lock-in detection technique (lower panel). In the measurements we used the following parameters: $V_{pump} = 90$ mV, $t_{pump} = 10$ μs, $V_{probe} = 5$ mV, $t_{probe} = 30$ μs. Typically, we applied 10–14 pump-probe pulse pairs per cycle. The repetition rate of cycle A and B was at the lock-in frequency of 89 Hz. (b) Pump-probe measurement on single Fe atom on MgO/Ag(001) at $B_z^{ext} = 1$ T, $B_\parallel^{ext} = -1.5$ T. Set point before opening the feedback loop was $V_{DC} = 10$ mV, $I = 100$ pA. An exponential fit yields a lifetime $T_1 = 34.6 \pm 2.5$ μs. The pump-probe signal was normalized to the background signal. (c) Extracted $T_1$ for different $B_\parallel^{ext}$ at $B_z^{ext} = 0.5$ T. Dotted line is a linear fit to the data. For all measurements the set point before opening the feedback loop was $V_{DC} = 10$ mV, $I = 50$ pA.

Besides the external magnetic field, the lifetime $T_1$ strongly depends on the tip-sample distance as well. Smaller tip-sample distances, i.e. higher conductances, lead to a strong decrease of $T_1$, because of the additional inelastic scattering with tip electrons[28]. From pump-probe measurements at different conductances ($V_{\text{DC}} = 10$ mV, $I = 50, 100, 150$ pA) at $B_z^{\text{ext}} = 1$ T and $B_\parallel^{\text{ext}} = -1.5$ T we estimated a ~3.5-fold decrease of $T_1$ for an increase of the setpoint current by 50 pA. This relation has been used to map the lifetime $T_1$ to our different conductance regimes in ESR.

## 4. Peak height intensity

In this section, we discuss the dependence of the ESR peak intensity $I_{\text{peak}}$ on the out-of-plane magnetic field $B_z^{\text{ext}}$ in greater detail. Assuming only one ground ($P_0$) and one excited state population ($P_1$), the probability for the state populations are given by $P_0 + P_1 = 1$. Furthermore, neglecting the influence of spin-pumping, a thermal Boltzmann distribution is given for the ground state and the population can be written as

$$\left[P_0 - \frac{1}{2}\right] = [1 + \exp(-hf_0/k_B T)]^{-1} - \frac{1}{2} \approx \frac{hf_0}{4k_B T} \tag{S1}$$

Here, $h$ is the Planck constant, $k_B$ is the Boltzmann constant and $T$ is the temperature. While the last equality is only valid for $hf_0 \ll k_B T$, it allow us to show the almost linear relation between $I_{\text{peak}}$ and $B_z^{\text{ext}}$. This is revealed by additionally making use of Eq. (1) and (2) in the main text, to obtain

$$I_{\text{peak}} = I \cdot 2\eta \cdot \left[P_0 - \frac{1}{2}\right] \cdot \Phi(\Omega) = \frac{\mu_{\text{Fe}} \cdot I \cdot \eta \cdot \Phi(\Omega)}{k_B T} \cdot \left[B_z^{\text{ext}} + B_z^{\text{tip}}\right] \tag{S2}$$

For fitting the data in Fig. 2C we employed Eq. (S2) without the approximation made in Eq. (S1). We used treated the prefactor $2\eta \cdot \Phi(\Omega)$ and the offset from the tip field as fitting parameters. Moreover, we used the experimental temperature of $T = 900$ mK.

## 5. Estimation of the Rabi rate

This section follows the discussion given in the supplement of Ref. (10). The ESR peak height can be expressed as[10]

$$I_{\text{peak}} = I_{\text{sat}} \cdot \Phi(\Omega) = I_{\text{sat}} \frac{\Omega^2 T_1 T_2}{1 + \Omega^2 T_1 T_2} \tag{S3}$$

where $T_1$ and $T_2$ are the energy relaxation and phase coherence time, respectively. This relation is directly obtained from the steady-state solution of the Bloch equation. Since we find experimentally $\Omega \propto V_{RF}$[7,10], the last term in Eq. (S3) can be rewritten as

$$\frac{\Omega^2 T_1 T_2}{1+\Omega^2 T_1 T_2} = \frac{(V_{RF}/V_{1/2})^2}{1+(V_{RF}/V_{1/2})^2} \tag{S4}$$

where $V_{1/2}$ is the half-saturation voltage[10]. The ESR amplitude increases with increasing RF power, until the system is fully driven into saturation, which is characterized by the saturation current of the ESR signal, which is given by $I_{sat} = \lim_{V_{RF} \to \infty} I_{peak}$[10]. The parameter $V_{1/2}$ is defined as the RF voltage that is required to achieve half of the saturation current, i.e. $I_{peak} = I_{sat}/2$ when $V_{RF} = V_{1/2}$, and therefore it defines the efficiency of driving the spin by the applied RF power. $V_{1/2}$ can be extracted by fitting the power dependence of the ESR peak amplitude using Eq. (S4). Extracted values of $V_{1/2}$ as a function of the in-plane magnetic field $B_\parallel^{ext}$ are shown in Fig. S4a for two different SP tips. As mentioned in the main text, we do not find any significant correlation between $V_{1/2}$ and the in-plane magnetic field, but a strong dependence of $V_{1/2}$ on the tunnel conductance $\sigma$.

From Eq. (S4), it follows by comparison that

$$\frac{\Omega}{V_{RF}} = \frac{1}{V_{1/2}\sqrt{T_1 T_2}} \tag{S5}$$

In order to extract the normalized Rabi rates $\Omega/V_{RF}$, we estimated $T_1$ and $T_2$ for our different ESR and conductance regimes from values found in other studies of the same spin system[7,10,28]. Note that each tunneling electron interacting with the Fe spin has a certain probability ($P_{T_2}$) to cause decoherence[10]. Consequently, the phase coherence time $T_2$ strongly decreases with increasing tunneling current following an inverse proportionality relation, i.e. $T_2 = e/(P_{T_2} I)$, where $e$ is charge of an electron and $I$ is tunnelling current[10]. We estimated $T_2$ using this relation and by assuming decoherence probability $P_{T_2} = 1$ for the sake of simplicity ($P_{T_2} = 0.7$ in Ref. [10]).

The energy relaxation time $T_1$ also depends on several experimental factors. This includes in particular the tunneling current and applied magnetic fields ($B_\parallel^{ext}$ and $B_z^{ext}$), as discussed in section 3. Therefore, the $T_1$ values measured using the pump probe method were first rescaled and extrapolated to match our ESR conditions of $I$, $B_z^{ext}$ and $B_\parallel^{ext}$ (see section 3). These $T_1$ values needed further rescaling to account for the effects of finite bias voltage which is always present during our ESR measurements. Note that the tunnel current significantly decreases $T_1$ because finite bias voltage present during ESR measurements readily allows the tunneling electrons to cause inelastic excitation of the Fe spin over the anisotropy barrier. However, it has only negligible influence on $T_2$[10] because the scattering electrons decohere the spin nearly independently of the electron's energy. Using values of previous $T_1$ measurements[7,10,28], we have estimated an expected reduction of $T_1$ by a factor of 3, 10 and 35 for datasets acquired at 8, 20 and 60 mV of DC bias voltage, respectively. These scaling factors were taken into account for the final estimations of $T_1$ in the three conductance regimes shown in Fig. 3a in the main text. The estimated $T_1$ values for different ESR conditions are listed in Table 1.

With these values of $V_{1/2}$, $T_1$ and $T_2$, we calculated the normalized Rabi rates following Eq. (5) for different conductance regimes. The results are shown in Fig. S4b for two different SP tips, including the one presented in Fig. 3b of the main text and different $B_z^{ext}$, $B_\parallel^{ext}$ and $\sigma$.

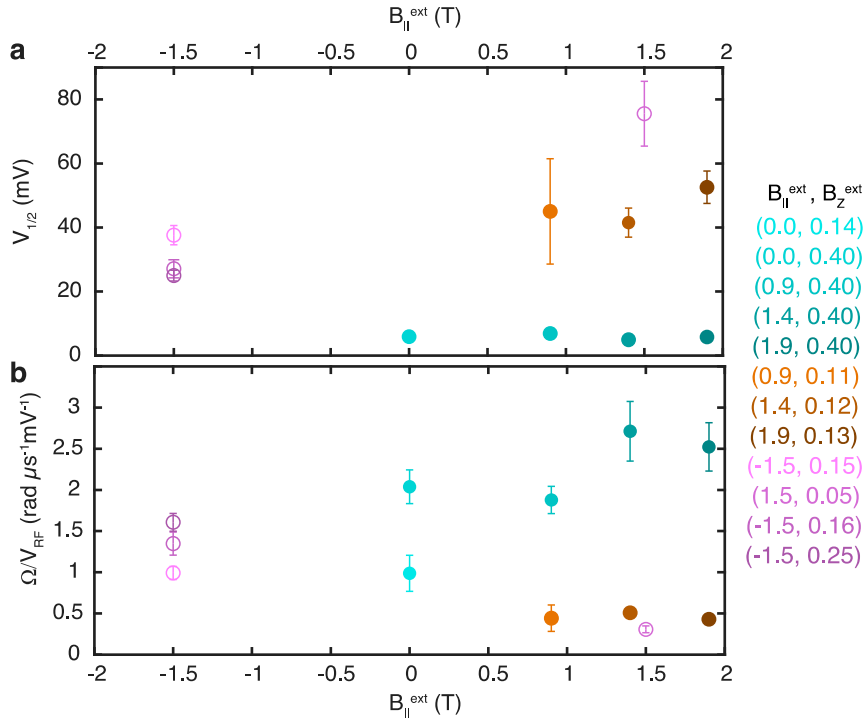

**Figure S4.** Power dependence and estimation of the Rabi rates. (a) Half-saturation voltage $V_{1/2}$ extracted by fitting the power dependence as shown in Fig 3b of the main text, as a function of in-plane magnetic field $B_\parallel^{ext}$. Error bars indicate standard error from the fit. (b) Normalized Rabi rates estimated using Eq. (S5), as a function of in-plane magnetic field $B_\parallel^{ext}$. Error bars represent the uncertainty estimated using propagation of error. For all panels, filled and open circles represent measurements with tip#1 and tip#2 respectively.

**Table 1: Estimated values of T₁ times at different ESR conditions**

| V | I | $B_{\parallel}^{ext}$ | $B_z^{ext}$ | $T_1$ estimated from $B_{\parallel}^{ext}$ dependence (See sec 3: pump probe spectroscopy) | Corrected $T_1$ considering applied $B_z^{ext}$ (See sec 3: pump probe spectroscopy) | Corrected $T_1$ considering tunnel current dependence | Corrected $T_1$ considering finite bias voltage |
|---|---|---|---|---|---|---|---|
| (mV) | (pA) | (T) | (T) | (µs) | (µs) | (µs) | (µs) |
| 8 | 105 | 0.0 | 0.40 | 52.0 | 47.4 | 13.5 | 4.5 |
| 8 | 104.5 | 0.9 | 0.40 | 45.5 | 40.9 | 11.7 | 3.9 |
| 8 | 103 | 1.4 | 0.40 | 41.9 | 37.3 | 10.7 | 3.6 |
| 8 | 110 | 1.9 | 0.40 | 38.3 | 33.7 | 9.6 | 3.2 |
| 60 | 50 | 0.9 | 0.11 | 45.5 | 27.6 | 27.6 | 0.8 |
| 60 | 50 | 1.4 | 0.12 | 41.9 | 24.6 | 24.6 | 0.7 |
| 60 | 50 | 1.9 | 0.13 | 38.3 | 21.4 | 21.4 | 0.6 |
| 60 | 50 | -1.5 | 0.15 | 24.00 | 7.9 | 7.9 | 0.2 |
| 60 | 50 | 1.5 | 0.05 | 41.2 | 20.4 | 20.4 | 0.6 |
| 60 | 50 | -1.5 | 0.16 | 24.00 | 8.2 | 8.2 | 0.2 |
| 20 | 92 | -1.5 | 0.25 | 23.95 | 12.5 | 3.6 | 0.4 |

## 6. Evaluation of the tip magnetic field as a function of conductance

We extracted the tip field $B_z^{tip}(\sigma)$ for different conductances by measuring $f_0(\sigma)$ in a series of ESR spectra. We varied the conductance $\sigma$ by changing the setpoint tunneling current $I$ and keeping the bias voltage $V_{DC}$ fixed. Using Eq. (1) in the main text then allowed us to extract $B_z^{tip}(\sigma)$ from $f_0(\sigma)$. We performed these measurements for a variety of different external vector magnetic fields ($B_{\parallel}^{ext}$ and $B_z^{ext}$). The results are summarized in Fig. S5. For all datasets the extracted tip field is linearly increasing with conductance in the measured range as expected for exchange interaction or magnetic dipolar interaction in a narrow range[25]. In the main text, we summarized the tip field for each dataset (given pair of $B_\parallel^{ext}$ and $B_z^{ext}$, indicated by color) by only showing the mean values of tip fields with respect to the mean conductance. Note that not all measurements extrapolate to zero tip-field for zero conductance. This is likely caused by a non-linear contribution of $B_z^{tip}(\sigma)$. Such a non-linearity is expected for either magnetic-dipolar interaction in a larger range of conductance or different scaling factors of exchange interaction and the exponential tunnel current dependence[9, 25].

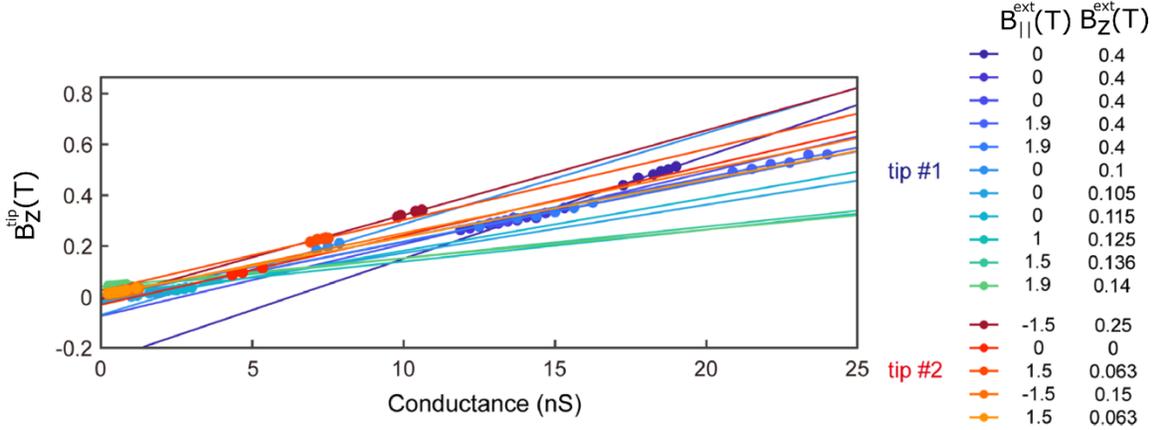

**Figure S5.** Tip fields as a function of tip-to-sample conductance at different external fields. Solid lines indicate linear fits to the data.

## 7. Relation between Rabi rate and tip field

In the main text we relate the tip field to the Rabi rate $\Omega \propto B_z^{tip}$. In this section, we derive this proportionality based on previous works[14,26]. Here, the Rabi rate was derived to be[14]

$$\frac{\Omega}{V_{RF}} = \frac{2e}{\hbar} \cdot \frac{1}{k \cdot d} \cdot \mathcal{F} \tag{S6}$$

where $k$ is the spring constant of the Fe atom on the MgO surface and $d$ is the decay constant of the electric field. Moreover, $\mathcal{F}$ is defined as the Rabi force, that in case of magnetic exchange interaction is given as

$$\mathcal{F}_J = \frac{\partial J(z)}{\partial z} \langle \vec{S}_{tip} \rangle \langle 0|\vec{S}|1\rangle = g\mu_B \frac{\partial \vec{B}_{tip}(z)}{\partial z} \langle 0|\vec{S}|1\rangle \tag{S7}$$

$\langle \vec{S}_{\text{tip}} \rangle$ is the statistical average of the tip-spin. In the second part, we have introduced the exchange tip field $\vec{B}_{\text{tip}}(z) = J(z)\langle \vec{S}_{\text{tip}} \rangle / g\mu_B$. The exchange interaction between tip and sample is in return often modelled as $J(z) = J_0 e^{-z/l}$[24-26], where $l$ is the decay length of the exchange interaction. Inserting this into Eq. (S7) yields

$$\mathcal{F}_J = \frac{\partial J(z)}{\partial z} \langle \vec{S}_{\text{tip}} \rangle \langle 0|\vec{S}|1\rangle = g\mu_B \frac{\partial \vec{B}_{\text{tip}}(z)}{\partial z} \langle 0|\vec{S}|1\rangle \qquad (S8)$$

And subsequently into Eq. (S6)

$$\frac{\Omega}{V_{RF}} = -\frac{2e}{\hbar} \cdot \frac{1}{k \cdot d} \frac{g\mu_B}{l} \vec{B}_{\text{tip}}(z) \langle 0|\vec{S}|1\rangle \propto \vec{B}_{\text{tip}}(z) \propto B_z^{\text{tip}} \qquad (S9)$$

This manifests the relation discussed in the context of Fig. 3. This proportionality holds for magnetic dipolar coupling in a similar way[14], though at these tunnel conductances magnetic exchange interaction usually dominates[25].

## 8. Evaluation and supplemental data on the ESR without macroscopic field

Here, we discuss the evaluation of the data shown in Fig. 4 in the manuscript, taken at $B_z^{\text{ext}} = B_\parallel^{\text{ext}} = 0$ T. In the main text we showed ESR measurements by sweeps of conductance $\sigma$. These sweeps of conductance were realized by sweeping the setpoint tunneling current $I$ under constant bias voltage conditions $V_{\text{DC}}$. In Fig. S6a we show the raw data revealing a nonlinear background signal likely stemming from non-linearities in the conductance as the tip approaches the atom. In order to eliminate this background, we subtracted a rescaled spectrum taken at 1 GHz, where the ESR peak is out of the sweep range. The resulting ESR spectra are shown in Fig. S6b, which are also displayed in Fig. 4b in the main text. Conversion of the tunnel current into conductance was done by $\sigma = I/V_{\text{DC}}$ with $V_{\text{DC}} = 30$ mV. In order to translate the swept tunneling current into an effective tip magnetic field, we measured the shift of the ESR peak under a finite external magnetic field of $B_z^{\text{ext}} = 50$ mT. We determined a shift in $I$ of $(58.7 \pm 2.1)$ pA (Fig. S7) leading to

$$\frac{\Delta B_z}{I(0 \text{ mT}) - I(50 \text{ mT})} = \frac{50 \text{ mT}}{(58.7 \pm 2.1) \text{ pA}} = (0.85 \pm 0.03) \frac{\text{mT}}{\text{pA}}$$

Using this conversion factor we translated the tunneling current into a magnetic tip field $B_z^{tip}$ (See upper x-scale in Fig. 4b in the manuscript) assuming a full linear scaling across the whole current range. We find evidence that this is a valid assumption from the almost perfect linear evolution of the resonant current (resonant tip magnetic field $B_z^{ext}$) as a function of frequency $f$ shown in Fig. S6c. The resonant current is the current for which ESR is achieved in Fig. S6b.

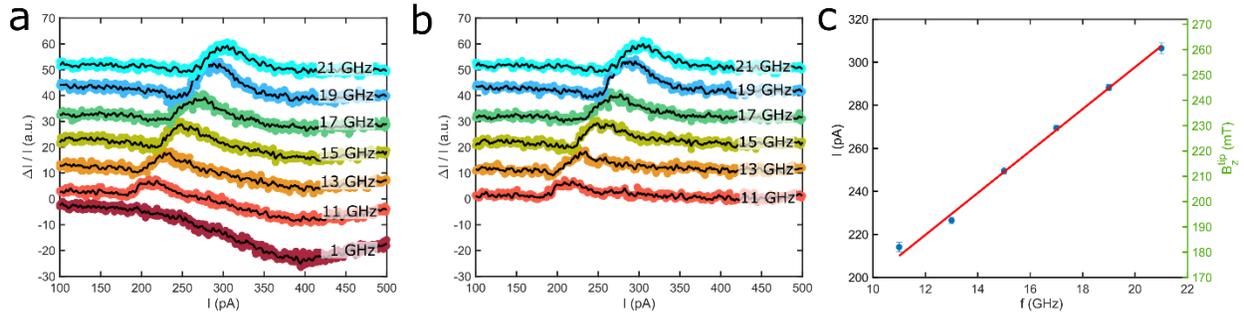

**Figure S6.** Current-sweep ESR spectra for different frequencies. (a) Raw data of zero-field ESR taken at different frequencies. Colored dots show two current sweeps and thin black line shows their average with a Gaussian filter applied. (b) ESR spectra as in (a) but with subtracted background signal**.** To subtract the background signal, the ESR sweep at 1 GHz [see panel (a)] was rescaled and subtracted. ($B_z^{ext} = B_{\parallel}^{ext} = 0$ T, $V_{DC} = 50$ mV, $f = 19$ GHz, $T = 0.4$ K). **c,** Evaluated resonant currents in (b) as a function of frequency $f$. Their position was determined by fitting a Fano lineshape to it. On the right the respective tip field (Obtained from Fig. S7, see text) is additionally plotted. The slope of this curve yields a Fe magnetic moment of $\mu_{Fe} = (4.29 \pm 0.79)\,\mu_B$, comparable to the one found in the main text.

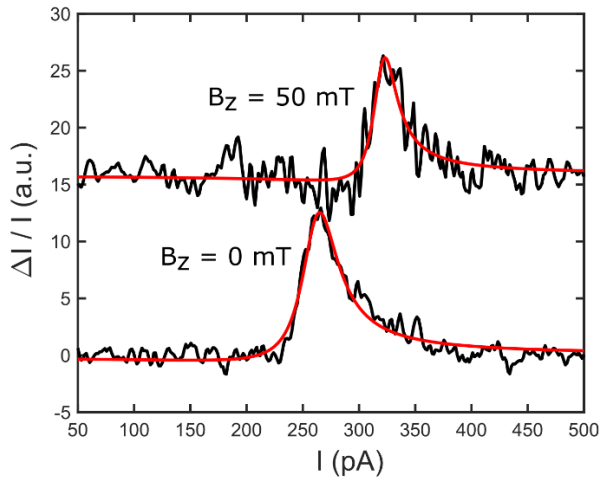

**Figure S7.** Shift of the ESR peak with a small external magnetic field $B_z^{\text{ext}} = 50$ mT. The resonance shifts $I$ by $(58.7 \pm 2.1)$ pA. The red lines are Fano lineshape fits to the data. ($V_{\text{DC}} = 50$ mV, $f = 19$ GHz, $T = 0.4$ K, $V_{\text{RF}} = 25$ mV)

REFERENCES

1. Balasubramanian, G.; Chan, I. Y.; Kolesov, R.; Al-Hmoud, M.; Tisler, J.; Shin, C.; Kim, C.; Wojcik, A.; Hemmer, P. R.; Krueger, A.; Hanke, T.; Leitenstorfer, A.; Bratschitsch, R.; Jelezko, F.; Wrachtrup, J. Nanoscale imaging magnetometry with diamond spins under ambient conditions. *Nature* **2008**, 455, 648-651.

2. Rugar, D.; Budakian, R.; Mamin, H.J.; Chui, B.W. Single spin detection by magnetic resonance force microscopy. *Nature* **2004**, 430, 329-332.

3. Degen, C.L.; Poggio, M.; Mamin, H.J.; Rettner, C.T.; Rugar, D.; Nanoscale magnetic resonance imaging. *PNAS* **2009**, 106, 1313-1317.

4. Gross, I.; Akhtar, W.; Garcia, V.; Martínez, L.J.; Chouaieb, S.; Garcia, K.; Carrétéro, C.; Barthélémy, A.; Appel, P.; Maletinsky, P.; Kim, J.V.; Chauleau, J. Y.; Jaouen, N. ; Viret, M.; Bibes, M.; Fusil, S.; Jacques, V. Real-space imaging of non-collinear antiferromagnetic order with a single-spin magnetometer. *Nature* **549**, 252-256 (2017).

5. Casola, F.; van der Sar, T.; Yacoby, A.; Probing condensed matter physics with magnetometry based on nitrogen-vacancy centres in diamond. *Nat. Rev. Mater.* **2018**, 3, 17088.

6. Thiel, L.; Wang, Z.; Tschudin, M.A.; Rohner, D.; Gutiérrez-Lezama, I.; Ubrig, N.; Gibertini, M.; Giannini, E.; Morpurgo, A.F.; Maletinsky, P.; Probing magnetism in 2D materials at the nanoscale with single spin microscopy. *Science* **2019,** 364, 973-976.

7. Baumann, S.; Paul, W.; Choi, T.; Lutz, C.P.; Ardavan, A.; Heinrich, A.J. Electron paramagnetic resonance of individual atoms on a surface. *Science* **2015**, 350, 417-420.

8. Choi, T.; Paul, W.; Rolf-Pissarczyk, S.; Macdonald, A.J.; Natterer, F.D.; Yang, K.; Willke, P.; Lutz, C.P.; Heinrich, A.J. Atomic-scale sensing of the magnetic dipolar field from single atoms. *Nature Nano.* **2017**, 12(5), 420-424.

9. Yang, K.; Bae, Y.; Paul, W.; Natterer, F.D.; Willke, P.; Lado, J.L.; Ferrón, A.; Choi, T.; Fernández-Rossier, J.; Heinrich, A.J.; Lutz, C.P. Engineering the eigenstates of coupled spin-1/2 atoms on a surface. *Phys. Rev. Lett.* **2017**, 119, 227206.

10. Willke, P.; Paul, W.; Natterer, F.D.; Yang, K.; Bae, Y.; Choi, T.; Fernández-Rossier, J.; Heinrich, A.J.; Lutz, C.P. Probing quantum coherence in single-atom electron spin resonance. *Science Adv.* **2018**, 4(2), eaaq1543.

11. Bae, Y.; Yang, K.; Willke, P.; Choi, T.; Heinrich, A.J.; Lutz, C.P. Enhanced quantum coherence in exchange coupled spins via singlet-triplet transitions. *Science Adv.* **2018**, 4(11), eaau4159.

12. Willke, P.; Bae, Y.; Yang, K.; Lado, J.L.; Ferrón, A.; Choi, T.; Ardavan, A.; Fernández-Rossier, J.; Heinrich, A.J.; Lutz, C.P. Hyperfine interaction of individual atoms on a surface. *Science* **2018**, 362(6412), 336-339.

13. Yang, K.; Willke, P.; Bae, Y.; Ferrón, A.; Lado, J.L.; Ardavan, A.; Fernández-Rossier, J.; Heinrich, A.J.; Lutz, C.P. Electrically controlled nuclear polarization of individual atoms. *Nature Nano*. **2018**, 13(12), 1120-1125.

14. Lado, J.L.; Ferrón, A.; Fernández-Rossier, J. Exchange mechanism for electron paramagnetic resonance of individual adatoms. *Phys. Rev. B* **2017**, 96(20), 205420.

15. Shakirov, A.M.; Rubtsov, A.N.; and Ribeiro, P. Spin transfer torque induced paramagnetic resonance. *Phys. Rev. B* **2019**, 99(5), 054434.

16. Berggren, P.; Fransson, J. Electron paramagnetic resonance of single magnetic moment on a surface. *Scientific reports* **2016**, 6, 25584.

17. Shavit, G.; Horovitz, B.; Goldstein, M. Generalized open quantum system approach for the electron paramagnetic resonance of magnetic atoms. *Phys. Rev. B* **2019**, 99(19), 195433.

18. Delgado, F.; Fernández-Rossier, J. Spin decoherence of magnetic atoms on surfaces. *Prog. Surf. Sci.* **2017**, 92(1), 40-82.

19. Ibañez-Azpiroz, J.; dos Santos Dias, M.; Blügel, S.; Lounis, S. Longitudinal and transverse spin relaxation times of magnetic single adatoms: An ab initio analysis. *Phys. Rev. B* **2017**, 96(14), 144410.

20. Gálvez, J.R.; Wolf, C.; Delgado, F.; Lorente, N. Cotunneling mechanism for all-electrical electron spin resonance of single adsorbed atoms. *Phys. Rev. B* **2019**, 100(3), 035411.

21. Natterer, F.D.; Patthey, F.; Bilgeri, T.; Forrester, P.R.; Weiss, N.; Brune, H. Upgrade of a low-temperature scanning tunneling microscope for electron-spin resonance. *Rev. Sci. Instrum.* **2019**, 90(1), 013706.

22. Seifert, T.S.; Kovarik, S.; Nistor, C.; Persichetti, L.; Stepanow, S.; Gambardella, P. Single-atom electron paramagnetic resonance in a scanning tunneling microscope driven by a radiofrequency antenna at 4 K. *arXiv preprint* **2019**, arXiv:1908.03379.

23. Paul, W.; Baumann, S.; Lutz, C. P.; Heinrich, A. J. Generation of constant-amplitude radio-frequency sweeps at a tunnel junction for spin resonance STM. *Rev. Sci. Instrum.* **2016**, 87, 074703.

24. Yan, S.; Choi, D.J.; Burgess, J.A.; Rolf-Pissarczyk, S.; Loth, S. Control of quantum magnets by atomic exchange bias. *Nature Nano.* **2010**, 10(1), 40-45.

25. Willke, P.; Yang, K.; Bae, Y.; Heinrich, A. J.; Lutz, C. P. Magnetic Resonance Imaging of Single Atoms, *Nat. Phys.* **2019**.

26. Yang, K.; Paul, W.; Natterer, F.D.; Lado, J.L.; Bae, Y.; Willke, P.; Choi, T.; Ferrón, A.; Fernández-Rossier, J.; Heinrich, A.J.; Lutz, C.P. Tuning the Exchange Bias on a Single Atom from 1 mT to 10 T. *Phys. Rev. Lett.* **2019**, 122(22), 227203.

27. Baumann, S.; Donati, F.; Stepanow, S.; Rusponi, S.; Paul, W.; Gangopadhyay, S.; Rau, I.G.; Pacchioni, G.E., Gragnaniello, L., Pivetta, M.; Dreiser, J.; Piamonteze, C.; Lutz, C. P.; Macfarlane, R. M.; Jones, B. A.; Gambardella, P.; Heinrich, A. J.; Brune, H. Origin of perpendicular magnetic anisotropy and large orbital moment in Fe atoms on MgO. *Phys. Rev. Lett.* **2015**, 115(23), 237202.

28. Paul, W.; Yang, K.; Baumann, S.; Romming, N.; Choi, T.; Lutz, C.P.; Heinrich, A.J. Control of the millisecond spin lifetime of an electrically probed atom. *Nat. Phys.* **2017**, 13, 403-407.

29. Khajetoorians, A.A.; Baxevanis, B.; Hübner, C.; Schlenk, T.; Krause, S.; Wehling, T.O.; Lounis, S.; Lichtenstein, A.; Pfannkuche, D.; Wiebe, J.; Wiesendanger, R. Current-driven spin dynamics of artificially constructed quantum magnets. *Science* **2013**, 339(6115), 55-59.

30. Loth, S.; Baumann, S.; Lutz, C.P.; Eigler, D.M.; Heinrich, A.J. Bistability in atomic-scale antiferromagnets. *Science* **2012**, 335(6065),196-199.

31. Forrester, P.R.; Patthey, F.; Fernandes, E.; Sblendorio, D.P.; Brune, H.; Natterer, F.D. A quantum pathway to overcome the trilemma of magnetic data storage. *arXiv preprint* **2019**, arXiv:1903.00242.